# Convective Viscous Cahn-Hilliard/Allen-Cahn Equation: Exact Solutions

P.O. Mchedlov-Petrosyan[*], L.N. Davydov[†]

NSC "Kharkov Institute of Physics and Technology, Kharkiv 61108, Ukraine

*Abstract*

Recently the combination of the well-known Cahn-Hilliard and Allen-Cahn equations was used to describe surface processes, such as simultaneous adsorption/desorption and surface diffusion. In the present paper we have considered the one-dimensional version of the Cahn-Hilliard/Allen-Cahn equation complemented with convective and viscous terms. Exact solutions are obtained and the conditions of their existence as well as the influence of applied field and additional dissipation are discussed.

## 1. Introduction.

Recently the combination of the well-known Cahn-Hilliard [1-4] and Allen-Cahn [5] equations was used to describe surface processes, such as simultaneous adsorption/desorption and surface diffusion.

Now, to understand the meaning of different terms and modifications of these equations, we need to give some insight into the history of CH and AC equations. The Cahn-Hilliard equation [1-4] is now a well-established model in the theory of phase transitions as well in several other fields. The basic underlying idea of this model is that for inhomogeneous system, e.g. system undergoing a phase transition, the thermodynamic potential (e.g. free energy) should depend not only on the order parameter $u$, but on its gradient as well. The idea of such dependence was introduced already by Van der Waals [6] in his theory of capillarity. For the inhomogeneous system the local chemical potential $\mu$ is defined as variational derivative of the thermodynamic potential functional. If the thermodynamic potential is the simplest symmetric – quadratic – function of gradient this leads to the local chemical potential $\mu$ which depends on Laplacian, or for the one-dimensional case – on the second order derivative of the order parameter (concentration). The diffusional flux $J$ is proportional to the gradient of chemical potential $\nabla \mu$; the coefficient of proportionality is called mobility $M$ [7]. With such expression for the flux the diffusion equation instead of usual second order equation becomes a forth-order PDE for the order parameter $u$ (herein our notations differ from the notations in original papers):

$$\frac{\partial u}{\partial t'} = \nabla \left[ M \nabla \mu \right], \tag{1.1}$$

$$\mu = -\bar{\varepsilon}^2 \Delta u + f(u). \tag{1.2}$$

---

[*] peter.mchedlov@free.fr
[†] ldavydov@kipt.kharkov.ua

Here $\bar{\varepsilon}$ is usually presumed to be proportional to the capillarity length and $f(u) = \dfrac{d\Phi(u)}{du}$, where $\Phi(u)$ is homogeneous part of the thermodynamic potential. In the present communication we will take $f(u)$ in the form of the cubic polynomial (corresponding to the fourth-order polynomial for the homogeneous part of thermodynamic potential):

$$f(u) = (u - a_1)(u - a_2)(u - a_3). \tag{1.3}$$

By rescaling $u$ the coefficient at $u^3$ could be always scaled to one. The classic Cahn-Hilliard equation was introduced as early as 1958 [1-2]; the stationary solutions were considered, the linearized version was treated and corresponding instability of homogeneous state identified. However, intensive study of the fully nonlinear form of this equation started essentially later [8]. Now an impressive amount of work is done on nonlinear Cahn-Hilliard equation, as well as on its numerous modifications, see [3-4]. An important modification was done by Novick-Cohen [9]. Taking into account the dissipation effects which are neglected in the derivation of the classic Cahn-Hilliard equation, she introduced the ***viscous*** Cahn-Hilliard (VCH) equation

$$\frac{\partial u}{\partial t'} = \nabla \left[ M \nabla \left( \mu + \bar{\eta} \frac{\partial u}{\partial t'} \right) \right], \tag{1.4}$$

where the coefficient $\bar{\eta}$ is called viscosity. It was also noticed that VCH equation could be derived as a certain limit of the classic Phase-Field model [10].

Later several authors considered the nonlinear **convective** Cahn-Hilliard equation (CCH) in one space dimension [11-13],

$$\frac{\partial u}{\partial t'} - \bar{\alpha} u \frac{\partial u}{\partial x'} = \frac{\partial}{\partial x'}\left(\frac{\partial \mu}{\partial x'}\right). \tag{1.5}$$

Leung [11] proposed this equation as a continual description of lattice gas phase separation under the influence of an external field. Similarly, Emmott and Bray [13] proposed this equation as a model for the spinodal decomposition of a binary alloy in an external field *E*. As they noticed, if the mobility *M* [7] is independent of the order parameter (concentration), the term involving *E* will drop out of the dynamics. To get nontrivial results, they presumed the simplest possible symmetric dependence of mobility on the order parameter, viz. $M \sim 1 - ru^2$. Then, they obtained the Burgers-type convection term in equation (1.5) with the coefficient $\bar{\alpha} = 2rE$. Thus, the sign of $\bar{\alpha}$ depends both on the direction of the field and on the sign of $r$. Witelski [12] introduced the equation (1.5) as a generalization of the classic Cahn–Hilliard equation or as a generalization of the Kuramoto–Sivashinsky equation [14-15] by including a nonlinear diffusion term. In [11-13] and [16-17] several approximate solutions and only two exact static kink and anti-kink solutions were obtained. The 'coarsening' of domains separated by kinks and anti-kinks was also discussed. To study the joint effects of nonlinear convection and viscosity, Witelski [18] introduced the convective-viscous-Cahn–Hilliard equation (CVCHE) with a general symmetric double-well potential $\Phi(u)$:

$$\frac{\partial u}{\partial t'} - \bar{\alpha} u \frac{\partial u}{\partial x'} = \frac{\partial}{\partial x'}\left[M \frac{\partial}{\partial x'}\left(\mu + \bar{\eta}\frac{\partial u}{\partial t'}\right)\right], \tag{1.6}$$

$$\mu = -\bar{\varepsilon}^2 \frac{\partial^2 u}{\partial x'^2} + \frac{d\Phi(u)}{du}. \tag{1.7}$$

It is worth noting that all results, including the stability of solutions, were obtained without specifying a particular functional form of the potential. Thus, they are valid both for the polynomial and logarithmic [3-4] potential. Also, with an additional constraint imposed on nonlinearity and viscosity, the approximate travelling-wave solutions were obtained. In [19] for equation (1.6) with polynomial potential, see (1.3), several exact single- and two-wave solutions were obtained.

Evidently, the Cahn-Hilliard equation describes the evolution of the locally conserved order parameter. For the evolution of the non-conserved order parameter Allen and Cahn [5] proposed the equation

$$\frac{\partial u}{\partial t'} = -\zeta\left[-\bar{\varepsilon}^2 \Delta u + f(u)\right], \qquad (1.8)$$

where $\zeta$ is a phenomenological constant. I.e., the change of the order parameter is proportional to the deviation of the local chemical potential (1.2) of an inhomogeneous system from its equilibrium value.

On the other hand, the catalysis and deposition from liquid and gaseous phase involve, besides the overall mass transport, a manifold of processes at the surface itself, e.g. attachment/detachment of atoms, surface diffusion, chemical reactions, etc. Traditionally these surface processes were modeled by continuum-type reaction-diffusion models, see e.g. [20]. Such approach usually neglects detailed interaction between the atoms. Recently a different model was developed [21] and studied in the series of papers [22-26]. On the microscopic level the bases of this approach is the modeling of surface processes by employing dynamic Ising type systems. Then the coarse-grained, mesoscopic models of Ising systems were developed [27-30]. As the next step, to illustrate the effects of multiple mechanisms, the simplification of mesoscopic equation (which retains its fundamental structure) for the order parameter $u$ was introduced [21-22]:

$$\frac{\partial u}{\partial t'} = \nabla\left[M\nabla\mu\right] - \zeta\mu. \qquad (1.9)$$

This is, evidently, the combination of the Cahn-Hilliard and Allen-Cahn equations. Here $\zeta$ is the proportionality constant in the Allen-Cahn equation, $\mu$ is given by (1.2)-(1.3); below we presume for definiteness $a_1 > a_2 > a_3$. In the present paper combining (1.9) and (1.6) we will consider the one-dimensional version of the *Convective-Viscous Cahn-Hilliard/Allen-Cahn equation* (CVCH/AC):

$$\frac{\partial u}{\partial t'} - \bar{\alpha}u\frac{\partial u}{\partial x'} = \frac{\partial}{\partial x'}\left[M\frac{\partial}{\partial x'}\left(\mu + \bar{\eta}\frac{\partial u}{\partial t'}\right)\right] - \zeta\mu, \qquad (1.10)$$

$$\mu = -\bar{\varepsilon}^2 \frac{\partial^2 u}{\partial x'^2} + (u-a_1)(u-a_2)(u-a_3). \tag{1.11}$$

In [22] the long-time behavior of large clusters was considered, so in (1.9) the time was rescaled with $\bar{\varepsilon}^2$ and space with $\bar{\varepsilon}$. For the one-dimensional problem which is considered here there is no physical reason for such scaling, see (1.10). Below we presume the mobility $M$ to be constant. Introducing the non-dimensional time $t = t'\zeta$, non-dimensional coordinate $x = x'\sqrt{\frac{\zeta}{M}}$, and $\alpha = \frac{\bar{\alpha}}{\sqrt{\zeta M}}$, $\varepsilon^2 = \bar{\varepsilon}^2 \frac{\zeta}{M}$, $\eta = \bar{\eta}\zeta$ we rewrite (1.10)-(1.11) as

$$\frac{\partial u}{\partial t} - \alpha u \frac{\partial u}{\partial x} = \frac{\partial^2}{\partial x^2}\left[-\varepsilon^2 \frac{\partial^2 u}{\partial x^2} + (u-a_1)(u-a_2)(u-a_3) + \eta \frac{\partial u}{\partial t}\right] + $$
$$+ \varepsilon^2 \frac{\partial^2 u}{\partial x^2} - (u-a_1)(u-a_2)(u-a_3). \tag{1.12}$$

## 2. Travelling wave solution.

Looking for the travelling wave solution of (1.12) we introduce the travelling wave coordinate $z = x - vt$. This yields

$$\frac{d}{dz}\left\{vu + \frac{\alpha u^2}{2} + \frac{d}{dz}\left[-\varepsilon^2 \frac{d^2 u}{dz^2} + (u-a_1)(u-a_2)(u-a_3) - \eta v \frac{du}{dz} + \varepsilon^2 u\right]\right\} = $$
$$= (u-a_1)(u-a_2)(u-a_3). \tag{2.1}$$

As usually we call "kinks" the solutions with $\frac{du}{dz} > 0$, and "anti-kinks" – solutions with $\frac{du}{dz} < 0$. We look for the anti-kink like solution, which connects the stationary state $a_1$ at $z = -\infty$ with the stationary state $a_3$ at $z = +\infty$. The simplest possible *Ansatz* for the solution with this property will be

$$\frac{du}{dz} = \kappa(u-a_1)(u-a_3), \tag{2.2}$$

where $\kappa$ is presently an unknown positive constant. Then it is, evidently,

$$\frac{1}{\kappa}\frac{d}{dz}\left[\frac{1}{2}u^2 - a_2 u\right] = \left(u - a_1\right)\left(u - a_2\right)\left(u - a_3\right). \tag{2.3}$$

Substituting the latter expression into (2.1) for the right-hand side and integrating once we get

$$\frac{d}{dz}\left[-\varepsilon^2 \frac{d^2 u}{dz^2} + \left(u - a_1\right)\left(u - a_2\right)\left(u - a_3\right) - \eta v \frac{du}{dz} + \varepsilon^2 u\right] + \\ + \left(\alpha - \frac{1}{\kappa}\right)\frac{u^2}{2} + \left(v + \frac{a_2}{\kappa}\right)u = C_1, \tag{2.4}$$

where $C_1$ is an arbitrary constant. Regarding the *Ansatz* (2.2), for the latter equation to be satisfied the expression under the derivative should be proportional to $u$. I.e., for (2.2) to give the solution of (2.1), two equations should be satisfied:

$$\beta \frac{du}{dz} + \left(\alpha - \frac{1}{\kappa}\right)\frac{u^2}{2} + \left(v + \frac{a_2}{\kappa}\right)u = C_1, \tag{2.5}$$

$$-\varepsilon^2 \frac{d^2 u}{dz^2} + \left(u - a_1\right)\left(u - a_2\right)\left(u - a_3\right) - \eta v \frac{du}{dz} + \varepsilon^2 u = \beta u + C_2, \tag{2.6}$$

where $\beta$ and $C_2$ are constants. The expressions for the derivatives are easily written as:

$$\frac{du}{dz} = \kappa\left(u^2 - Xu + Y\right), \tag{2.7}$$

$$\frac{d^2 u}{dz^2} = \kappa^2\left[2u^3 - 3Xu^2 + \left(2Y + X^2\right)u - XY\right], \tag{2.8}$$

where we have denoted for brevity

$$X = a_1 + a_2; \; Y = a_1 a_2. \tag{2.9}$$

7Then Eqs. (2.5), (2.6) take the form

$$\left(\beta\kappa + \frac{\alpha}{2} - \frac{1}{2\kappa}\right)u^2 + \left(-\beta\kappa X + v + \frac{a_2}{\kappa}\right)u = C_1, \qquad (2.10)$$

$$\begin{aligned}&-\varepsilon^2\kappa^2\left[2u^3 - 3Xu^2 + \left(2Y + X^2\right)u\right] + \\ &+\left[u^3 - \left(X + a_2\right)u^2 + \left(Y + a_2 X\right)u\right] - \\ &-\eta v\kappa\left(u^2 - Xu\right) + \left(\varepsilon^2 - \beta\right)u = C_2.\end{aligned} \qquad (2.11)$$

These equations should be satisfied for arbitrary $u$. In (2.10) and (2.11) zero-power in $u$ terms are absorbed by the arbitrary constants $C_1$ and $C_2$. Rearranging the terms and equating coefficients at each power of $u$ to zero we finally obtain five constraints on the parameters,

$$2\varepsilon^2\kappa^2 = 1, \qquad (2.12)$$

$$\frac{1}{2}X - a_2 - \eta v\kappa = 0, \qquad (2.13)$$

$$-\frac{1}{2}X^2 + \left(a_2 + \eta v\kappa\right)X + \varepsilon^2 - \beta = 0, \qquad (2.14)$$

$$\beta\kappa + \frac{\alpha}{2} - \frac{1}{2\kappa} = 0, \qquad (2.15)$$

$$-\beta\kappa X + v + \frac{a_2}{\kappa} = 0. \qquad (2.16)$$

If the constraints (2.12)-(2.16) are satisfied, the corresponding solutions of the first-order equation (2.2) are simultaneously exact travelling-wave solutions of equation (1.12). Equation (2.2) is easily integrated, yielding

$$u = \frac{a_1 + a_3 \exp\left\{\kappa\left(a_1 - a_3\right)\left(z + \varphi\right)\right\}}{1 + \exp\left\{\kappa\left(a_1 - a_3\right)\left(z + \varphi\right)\right\}} \qquad (2.17)$$

where $\varphi$ is an arbitrary constant. It is natural to take position of the maximal value of the derivative $\dfrac{du}{dz}$ (where $\dfrac{d^2u}{dz^2}=0$), as $z=0$; then $\varphi=0$. The solution (2.2) could be rewritten in the form

$$u = \frac{a_1+a_3}{2} - \frac{a_1-a_3}{2} th\left[\frac{1}{2}\kappa\left(a_1-a_3\right)\left(x-vt\right)\right]. \qquad (2.18)$$

Now let us consider the constraints (2.12)-(2.16). From (2.13) and (2.14) it follows

$$\beta = \varepsilon^2. \qquad (2.19)$$

From (2.12), (2.15) and (2.19),

$$\alpha\kappa = 1 - 2\kappa^2\varepsilon^2 = 0. \qquad (2.20)$$

This means, that the exact constant velocity travelling-wave solution is possible only in the absence of applied field. From (2.13) and (2.16),

$$a_2 + \eta v\kappa = a_2 + v\kappa. \qquad (2.21)$$

I.e., necessary either $\eta=1$, or $v=0$. If $\eta=1$ the velocity of the transition front is

$$v = \varepsilon\sqrt{2}\left(\frac{a_1+a_3}{2} - a_2\right). \qquad (2.22)$$

## *3. Discussion*

In the present paper we have considered the one-dimensional version of the Convective-Viscous Cahn-Hilliard/Allen-Cahn equation (CVCH/AC) (1.10). The combination of classic Cahn-Hilliard and Allen-Cahn equations was introduced [21-22] to model the surface processes involving attachment/detachment of atoms and surface diffusion. Our aim was to explore the influence of additional factors, such as external field (modeled by the convective term), and additional dissipation (modeled by viscous term) on this model. It is well known that for the classic Cahn-Hilliard equation only exact static solutions are possible. Nevertheless, the introduction of the convective and viscous terms enables, for the balance of external field and dissipation, existence of exact one- and two-wave solutions [19].

However, for the CVCHAC the situation is essentially different. The very presence of external field makes the existence of exact constant-velocity travelling wave solution impossible (2.20). On the other hand, the existence of such solution is ensured by the balance between the additional dissipation in diffusion (i.e. in the "Cahn-Hilliard part") and "Allen-Cahn part", $\eta = 1$, or in the initial notations $\bar{\eta} = 1/\zeta$. Otherwise only exact static solutions (as for the classic Cahn-Hilliard equation) are possible. In the initial notation the dimensional velocity $V$ is

$$V = \bar{\varepsilon}\zeta\sqrt{2}\left(\frac{a_1 + a_3}{2} - a_2\right). \tag{3.1}$$

Remarkably, while the dependence of velocity on the values of stationary states is exactly the same, as for the well known solution of the diffusion equation with cubic nonlinearity, i.e. on the Allen-Cahn part, the coefficient depends on the capillarity length $\bar{\varepsilon}$, i.e. on the Cahn-Hilliard part. It evidently reminds the appearance of the surface tension in the expression for the velocity for the motion by mean curvature [22], though the "driving forces" are essentially different.

5. Allen, S.M. & Cahn, J.W. (1979) *A microscopic theory for antiphase boundary motion and its application to antiphase domain coarsening.* Acta Metallurgica **27**, pp. 1085-1095.

6. Van der Waals, J.D. (1979) *The thermodynamic theory of capillarity under the hypothesis of a continuous variation of* density (1893), Translated by J. S. Rowlinson. Journal of Statistical Physics **20**, pp.197-244.

7. De Groot S.R., Mazur P. (1984) *Non-Equilibrium Thermodynamics,* Dover.

8. Novick-Cohen, A. & Segel, L.A. (1984) *Nonlinear aspects of the Cahn–Hilliard equation.* Physica D**10**, pp.277-298.

9. Novick-Cohen, A. (1988) *On the viscous Cahn_Hilliard equation.* In: "Material Instabilities in Continuum Mechanics and Related Mathematical Problems" (J. M. Ball, Ed.), Oxford Univ. Press, Oxford, pp.329-342.

10. Bai, F., Elliott, C.M., Gardiner, A., Spence, A., & Stuart, A.M. (1995) *The viscous Cahn-Hilliard equation. I. Computations*, Nonlinearity **8**, 131-160.

11. Leung, K. (1990) *Theory on Morphological Instability in Driven systems.* Journal of Statistical Physics **61**, pp.345-364.

12. Witelski, T.P., (1995), *Shocks in Nonlinear Diffusion.* Appl.Math.Lett. **8**, pp.27-32.

13. Emmott C.L. & Bray, A.J. (1996) *Coarsening dynamics of a one-dimensional driven Cahn-Hilliard system,* Phys. Rev. E**54** (5), pp.4568-4575.

14. Kuramoto, Y. & Tsuzuki, T. (1976) *Persistent propagation of concentration waves in dissipative media far from thermal equilibrium.* Progr. Theoret. Phys. **55**, pp.356-369.

15. Sivashinsky, G. (1977*) Nonlinear analysis of hydrodynamic instability in laminar flames I. Derivation of basic equations.* Acta Astron. **4**, pp.1177-1206.

16. Watson, S. J., Otto, F., Rubinstein, B. Y. & Davis, S. H. (2002) *Coarsening dynamics for the convective Cahn-Hilliard equation.* Max-Plank-Institut fuer Mathematik in den Naturwissenschaften, Leipzig, Preprint no. 35.